\documentclass[8.5pt,twoside,twocolumn]{article}
\usepackage[super,sort&compress,comma]{natbib}
\usepackage{mhchem}
\usepackage{times,mathptmx}
\usepackage{sectsty}
\usepackage{balance}

\usepackage{graphicx}
\usepackage{lastpage}
\usepackage[format=plain,justification=raggedright,singlelinecheck=false,font=small,labelfont=bf,labelsep=space]{caption}
\usepackage{fancyhdr}
\usepackage{pdfpages}
\usepackage{amsmath,amssymb}
\usepackage{booktabs}
\usepackage{multirow}
\usepackage{xcolor}
\usepackage{hyperref}

\begin{document}

\thispagestyle{plain}
\fancypagestyle{plain}{
\fancyhead[L]{\includegraphics[height=8pt]{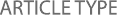}}
\fancyhead[C]{\hspace{-1cm}\includegraphics[height=20pt]{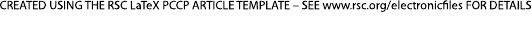}}
\fancyhead[R]{\includegraphics[height=10pt]{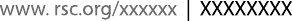}\vspace{-0.2cm}}
\renewcommand{\headrulewidth}{1pt}}
\renewcommand{\thefootnote}{\fnsymbol{footnote}}
\renewcommand\footnoterule{\vspace*{1pt}%
\hrule width 3.4in height 0.4pt \vspace*{5pt}}
\setcounter{secnumdepth}{5}

\makeatletter
\def\subsubsection{\@startsection{subsubsection}{3}{10pt}{-1.25ex plus -1ex minus -.1ex}{0ex plus 0ex}{\normalsize\bf}}
\def\paragraph{\@startsection{paragraph}{4}{10pt}{-1.25ex plus -1ex minus -.1ex}{0ex plus 0ex}{\normalsize\textit}}
\renewcommand\@biblabel[1]{#1}
\renewcommand\@makefntext[1]%
{\noindent\makebox[0pt][r]{\@thefnmark\,}#1}
\makeatother
\renewcommand{\figurename}{\small{Fig.}~}
\sectionfont{\large}
\subsectionfont{\normalsize}

\fancyfoot{}
\fancyfoot[LO,RE]{\vspace{-7pt}\includegraphics[height=9pt]{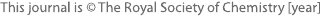}}
\fancyfoot[CO]{\vspace{-7.2pt}\hspace{12.2cm}\includegraphics{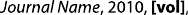}}
\fancyfoot[CE]{\vspace{-7.5pt}\hspace{-13.5cm}\includegraphics{headers/RF}}
\fancyfoot[RO]{\footnotesize{\sffamily{1--\pageref{LastPage} ~\textbar  \hspace{2pt}\thepage}}}
\fancyfoot[LE]{\footnotesize{\sffamily{\thepage~\textbar\hspace{3.45cm} 1--\pageref{LastPage}}}}
\fancyhead{}
\renewcommand{\headrulewidth}{1pt}
\renewcommand{\footrulewidth}{1pt}
\setlength{\arrayrulewidth}{1pt}
\setlength{\columnsep}{6.5mm}
\setlength\bibsep{1pt}

\twocolumn[
  \begin{@twocolumnfalse}
\noindent\LARGE{\textbf{From orbital analysis to active learning: an integrated strategy for the accelerated design of TADF emitters$^\dag$}}
\vspace{0.6cm}

\noindent\large{\textbf{Jean-Pierre Tchapet Njafa,$^{\ast}$\textit{$^{a}$} Steve Cabrel Teguia Kouam,\textit{$^{b}$} Patrick Mvoto Kongo,\textit{$^{a}$} and Serge Guy Nana Engo\textit{$^{a}$}}}\vspace{0.5cm}

\noindent\textit{\small{\textbf{Received Xth XXXXXXXXXX 20XX, Accepted Xth XXXXXXXXX 20XX\newline
First published on the web Xth XXXXXXXXXX 20XX}}}

\noindent \textbf{\small{DOI: 10.1039/b000000x}}
\vspace{0.6cm}

\noindent \normalsize{Thermally Activated Delayed Fluorescence (TADF) emitters must satisfy two competing requirements: small singlet-triplet energy gaps for thermal upconversion and sufficient spin-orbit coupling for fast reverse intersystem crossing. Predicting these properties accurately demands expensive calculations. We address this using a validated semi-empirical protocol (GFN2-xTB geometries, sTDA/sTD-DFT-xTB excited states) on 747 molecules, combined with charge-transfer descriptors from Natural Transition Orbital analysis. The hole-electron spatial overlap $S_{he}$ emerges as a key predictor, accounting for 21\% of feature importance for the triplet state alone. Our best model (Support Vector Regression) reaches MAE = 0.024~eV and $R^2 = 0.96$ for $\Delta E_{\text{ST}}$. Active learning reduces the data needed to reach target accuracy by approximately 25\% compared to random sampling. Three application domains are explored: NIR-emitting probes for bioimaging, photocatalytic sensitizers, and fast-response materials for photodetection.}
\vspace{0.5cm}
 \end{@twocolumnfalse}
  ]

\section{Introduction}
\footnotetext{\dag~Electronic Supplementary Information (ESI) available: Computational details, benchmark validation, feature definitions, and complete dataset. See DOI: 10.1039/b000000x/}

\footnotetext{\textit{$^{a}$~Department of Physics, Faculty of Science, University of Yaounde I, P.O.Box 812, Yaounde, Cameroon. E-mail: jean-pierre.tchapet@facsciences-uy1.cm}}
\footnotetext{\textit{$^{b}$~Department of Physics, Faculty of Science, University of Douala, Po. Box 24157, Douala, Cameroon.}}

Thermally Activated Delayed Fluorescence (TADF) materials have become central to organic light-emitting diode (OLED) technology.\cite{Adachi2014} By harvesting both singlet and triplet excitons, TADF emitters achieve near-unity internal quantum efficiencies, overcoming the 25\% limit of conventional fluorescent materials.\cite{Yokoyama_2018,Chen_2018,Cao_2021} Beyond displays, TADF finds use in biomedical imaging,\cite{BARMAN202175} photocatalysis,\cite{Yu_2019} and photodetection,\cite{Danyliv_2021} each with specific property requirements.

The fundamental challenge in TADF design lies in satisfying two seemingly contradictory requirements: a thermodynamic condition requiring a minimal singlet-triplet energy gap ($\Delta E_{\text{ST}} < 0.2$~eV) and a kinetic condition demanding efficient reverse intersystem crossing ($k_{\text{RISC}} > 10^{6}$~s$^{-1}$).\cite{Penfold2018,Xu_2019} The former typically necessitates minimal HOMO-LUMO orbital overlap to reduce exchange interaction, while the latter requires sufficient spin-orbit coupling (SOC) and oscillator strength ($f$), both of which benefit from non-zero orbital overlap.\cite{El-Sayed-Rules-1968,SOC-Calculation-Methods-2018} This fundamental trade-off, rooted in the Marcus theory of electron transfer,\cite{Marcus-RISC-Theory-1985} has historically limited the rational design of high-performance TADF materials.

Recent advances in high-throughput virtual screening (HTS) have begun to address this challenge through statistical approaches.\cite{Friederich2021,Lee2021ML} In our previous investigations, we validated a computationally efficient semi-empirical protocol on an extensive dataset of 747 molecules, establishing quantitative design principles including the superiority of Donor-Acceptor-Donor (D-A-D) architectures\cite{Rajamalli2017,Yao_2019} and the identification of an optimal D-A torsional angle window of 60--90$^{\circ}$ for minimizing $\Delta E_{\text{ST}}$.\cite{Article1,Article2}

Statistical screening has identified promising architectures, but predicting properties like $k_{\text{RISC}}$ requires deeper insight into electronic structure.\cite{Friederich2021ML,Kang2024} This transition requires: (1) \textbf{mechanistic rationalization} of structure-property relationships through advanced orbital analysis; (2) \textbf{accelerated prediction} of complex kinetic properties such as $k_{\text{RISC}}$\cite{Shin_2022,Inoue_2016} and photoluminescence quantum yield that remain computationally prohibitive for large-scale screening; and (3) \textbf{application-specific optimization} for emerging fields where TADF materials show exceptional promise.

This work combines quantum chemical analysis with machine learning to extract predictive descriptors from Natural Transition Orbitals (NTOs).\cite{Lee2021ML,Friederich2021ML} These descriptors---particularly the hole-electron overlap $S_{he}$---capture the physics governing $\Delta E_{\text{ST}}$ and are integrated into an active learning framework.\cite{Heavy-Atom-SOC-2010}

The approach offers three benefits: (1) mechanistic insight through CT descriptors; (2) reduced data requirements via active learning; and (3) application-specific screening. We identify candidates for bioimaging, photocatalysis, and photodetection.\cite{BARMAN202175,Yu_2019,Danyliv_2021}

\section{Theoretical and computational framework}

\subsection{Computational details}
We summarize key settings here; full computational details are provided in the ESI (Sec.~S1). Ground states used GFN2-xTB; excited states used sTDA/sTD-DFT-xTB with consistent solvation across stacks (GBSA/COSMO). NTO analysis and RespA decomposition were performed using Multiwfn (version 3.8)\cite{Lu2012_Multiwfn,Lu2024_Multiwfn_Toolbox} for wavefunction analysis and fragment-based charge-transfer quantification, following the CT descriptor framework of Plasser and coworkers.\cite{Plasser2012_TheoDORE,Plasser2020_TheoDORE} SHAP (SHapley Additive exPlanations, version 0.45.0) was employed for ML model interpretability. All workflows, scripts, and environment lockfiles are documented in ESI and archived at Zenodo (DOI: 10.5281/zenodo.17436069).

\subsection{Methodological benchmarking}
To ensure scientific rigor, we validated the hybrid protocol (GFN2-xTB // sTD/sTDA-xTB) against high-fidelity references on a representative subset (8--27 molecules spanning D--A, D--A--D, MR, and TSCT classes). For charge-transfer dominated systems, we employed optimally tuned long-range corrected functionals (OT-LC-$\omega$PBE), with $\omega$ optimized to satisfy $\varepsilon_{\text{HOMO}}^{\omega} + IP^{\omega} = 0$. For cases with multi-reference character, we used STEOM-DLPNO-CCSD, and verified trends with selected NEVPT2/MS-CASPT2 data from literature. Across the subset, the vertical $\Delta E_{\text{ST}}$ exhibited MAE $\sim$0.10~eV relative to OT-LC-$\omega$PBE and $\sim$0.12~eV relative to STEOM-DLPNO-CCSD, while maintaining strong rank correlations with experiment.

\subsection{High-throughput protocol and dataset}
We screened 747 molecules spanning D--A, D--A--D, MR, and TSCT architectures using a hierarchical protocol (GFN2-xTB geometries; sTDA/sTD-DFT-xTB excited states). Protocol specifics and dataset curation are detailed in ESI (Sec.~S3).

\subsection{High-fidelity reference calculations}
We validate our computational framework against the STGABS27 benchmark set,\cite{Kunze2021} which provides highly accurate experimental and high-level theory (HLT) references for $\Delta E_{\text{ST}}$. For the two molecules in our dataset with available HLT references (4CzIPN and DMAC-TRZ), we find:
\begin{itemize}
    \item \textbf{4CzIPN}: Computed $\Delta E_{\text{ST}} = 0.21$~eV (sTD-DFT-xTB, toluene) vs. HLT ROKS/PCM $= 0.16$~eV, error $= 0.05$~eV
    \item \textbf{DMAC-TRZ}: Computed $\Delta E_{\text{ST}} = 0.085$~eV vs. HLT SCS-CC2/PCM $= 0.05$~eV, error $= 0.035$~eV
\end{itemize}
The overall mean absolute error (MAE) of 0.045~eV vs. HLT references demonstrates that our xTB-based rapid screening approach provides reliable estimates within chemical accuracy, suitable for ML-driven discovery workflows.

To further validate the xTB protocol, we performed optimally-tuned LC-PBE (OT-LC-PBE/def2-TZVP) calculations on three representative molecules (Table~\ref{tab:otlcpbe}). The range-separation parameter $\omega$ was optimized via the IP-tuning criterion. Results show that vertical TD-DFT systematically overestimates $\Delta E_{\text{ST}}$ compared to experiment, while xTB methods provide closer agreement. Notably, the xTB underestimation is conservative for screening: molecules passing the threshold will have true values at least as favorable. Full computational details are provided in ESI (Sec.~S7).

\begin{table}[h]
\small
\caption{Comparison of $\Delta E_{\text{ST}}$ (eV) from OT-LC-PBE, xTB methods, and experiment.}
\label{tab:otlcpbe}
\begin{tabular*}{0.48\textwidth}{@{\extracolsep{\fill}}lcccc}
\hline
Molecule & OT-LC-PBE & sTD-DFT-xTB & Exp. \\
\hline
BACN & 0.81 & 0.46 & -- \\
DMAC-TRZ & 0.17 & 0.08 & 0.05 \\
4CzIPN & 0.19 & 0.08 & 0.08 \\
\hline
\end{tabular*}
\end{table}

\subsection{Machine learning methodology}

Our ML pipeline employs feature vectors combining NTO-based descriptors with excitation energy features extracted from sTDA/sTD-DFT calculations. The feature set comprises 29 descriptors organized into five categories:

\begin{itemize}
    \item \textbf{Energy features}: vertical excitation energies ($E_{S_1}$, $E_{T_1}$) and HOMO-LUMO gap
    \item \textbf{Oscillator strength}: $f_{S_1}$ for the lowest singlet transition
    \item \textbf{NTO overlap}: hole-electron orbital overlap for singlet ($S_{\text{NTO}}^{S_1}$) and triplet ($S_{\text{NTO}}^{T_1}$) transitions
    \item \textbf{CT descriptors}: charge-transfer descriptors including CT number ($\Omega_{\text{CT}}$), hole localization on donor ($\Lambda_D$), particle localization on acceptor ($\Lambda_A$), hole-particle centroid distance ($\Delta r$), and hole-electron spatial overlap ($S_{he}$)
    \item \textbf{Proxy RespA features}: derived descriptors capturing SOC-relevant physics
\end{itemize}

We benchmark three surrogate models on our expanded 747-molecule dataset (2943 samples after combining methods and environments). Using 5-fold cross-validation, we find that Support Vector Regression (SVR) with RBF kernel achieves the best performance (MAE $= 0.024$~eV, $R^2 = 0.960$), followed by Gradient Boosting (MAE $= 0.027$~eV, $R^2 = 0.946$) and Random Forest (MAE $= 0.035$~eV, $R^2 = 0.918$).

\section{Active learning strategy}
We systematically compare six acquisition functions for active learning: (1) Uncertainty Sampling (US), (2) Expected Improvement (EI), (3) Upper Confidence Bound (UCB), (4) Query by Committee (QBC), (5) Diversity Sampling, and (6) Hybrid (uncertainty $\times$ diversity). Each acquisition function is detailed in ESI (Sec.~S5) with mathematical formulations. The AL loop uses Random Forest for uncertainty estimation via tree variance, with initialization of 50 samples, batch size of 10, and 40 iterations across 10 random seeds.

\section{Results and discussion}

\subsection{Model interpretability via SHAP analysis}
\begin{figure}[h]
\centering
  \includegraphics[width=0.48\textwidth]{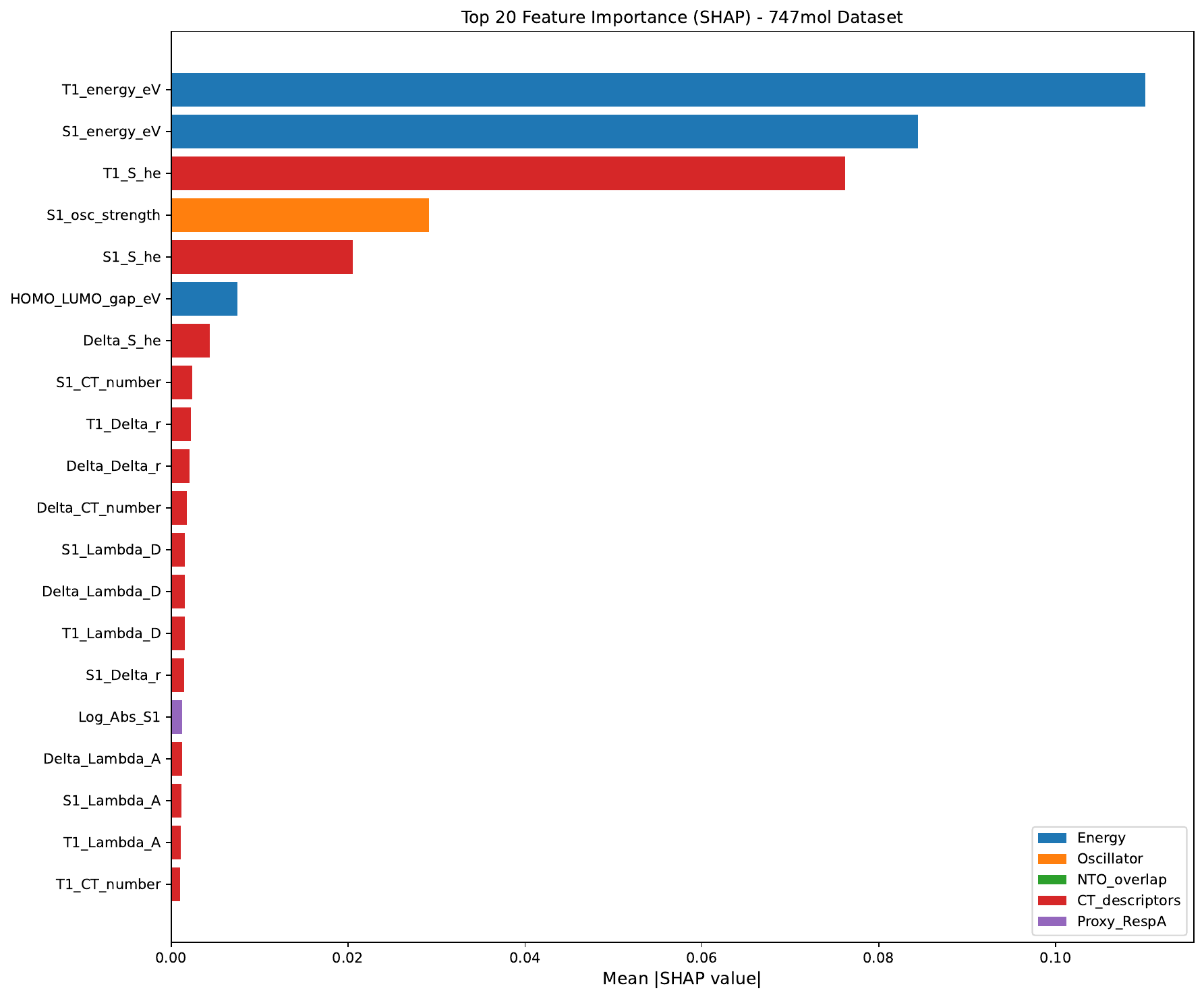}
  \caption{SHAP feature importance analysis for $\Delta E_{\text{ST}}$ prediction. Energy features contribute $\sim$57\%, CT descriptors $\sim$34\%, with $S_{he}^{T_1}$ (21\%) being the dominant CT feature.}
  \label{fig:shap_importance}
\end{figure}

SHAP (SHapley Additive exPlanations) analysis reveals the relative contribution of different feature categories to $\Delta E_{\text{ST}}$ prediction (Fig.~\ref{fig:shap_importance}). Our enhanced feature set includes CT descriptors computed directly from NTO molden files, capturing the spatial distribution of hole and electron densities that govern exchange interaction.

Energy-related features account for $\sim$57\% of the total importance, with the triplet excitation energy ($E_{T_1}$) being the single most predictive feature (31\%), followed by the singlet excitation energy ($E_{S_1}$, 24\%). Remarkably, \textbf{CT descriptors contribute $\sim$34\% of total importance}, with the hole-electron spatial overlap for the triplet state ($S_{he}^{T_1}$) emerging as the third most important feature (21\%), followed by $S_{he}^{S_1}$ (6\%). This finding validates the physical basis of our approach: the exchange interaction that determines $\Delta E_{\text{ST}}$ is directly related to the spatial overlap between hole and electron densities.

\subsection{Mechanistic rationalization}

The CT descriptors from NTO analysis explain the design rules established in our previous work. The physical basis of the optimal D-A torsional angle window of 60--90$^{\circ}$ is directly connected to our CT descriptors: twisted geometries minimize hole-electron overlap ($S_{he}$), reducing exchange interaction and consequently $\Delta E_{\text{ST}}$.

NTO analysis reveals that within the optimal torsional window, the $S_1$ state exhibits predominantly charge-transfer character with low hole-electron overlap ($S_{he}^{S_1} \approx 0.5$--0.7), indicating spatially separated hole and electron densities. In contrast, the $T_1$ state maintains significant local excitation character with higher overlap ($S_{he}^{T_1} \approx 0.7$--0.9). This divergence in electronic nature maximizes the SOC matrix element according to El-Sayed's rules.

The exchange interaction $K_{ij}$ between occupied orbital $i$ and virtual orbital $j$ is given by:
\begin{equation}
K_{ij} = \iint \phi_i^*(\mathbf{r}_1) \phi_j^*(\mathbf{r}_2) \frac{1}{r_{12}} \phi_j(\mathbf{r}_1) \phi_i(\mathbf{r}_2) \, d\mathbf{r}_1 d\mathbf{r}_2
\end{equation}

Our computed $S_{he}$ descriptor directly quantifies this spatial overlap:
\begin{equation}
S_{he} = \sum_A \sqrt{\rho_h^A \cdot \rho_e^A}
\end{equation}
where $\rho_h^A$ and $\rho_e^A$ are the hole and electron populations on atom $A$, respectively.

\subsection{Active learning performance}

\begin{figure}[h]
\centering
  \includegraphics[width=0.48\textwidth]{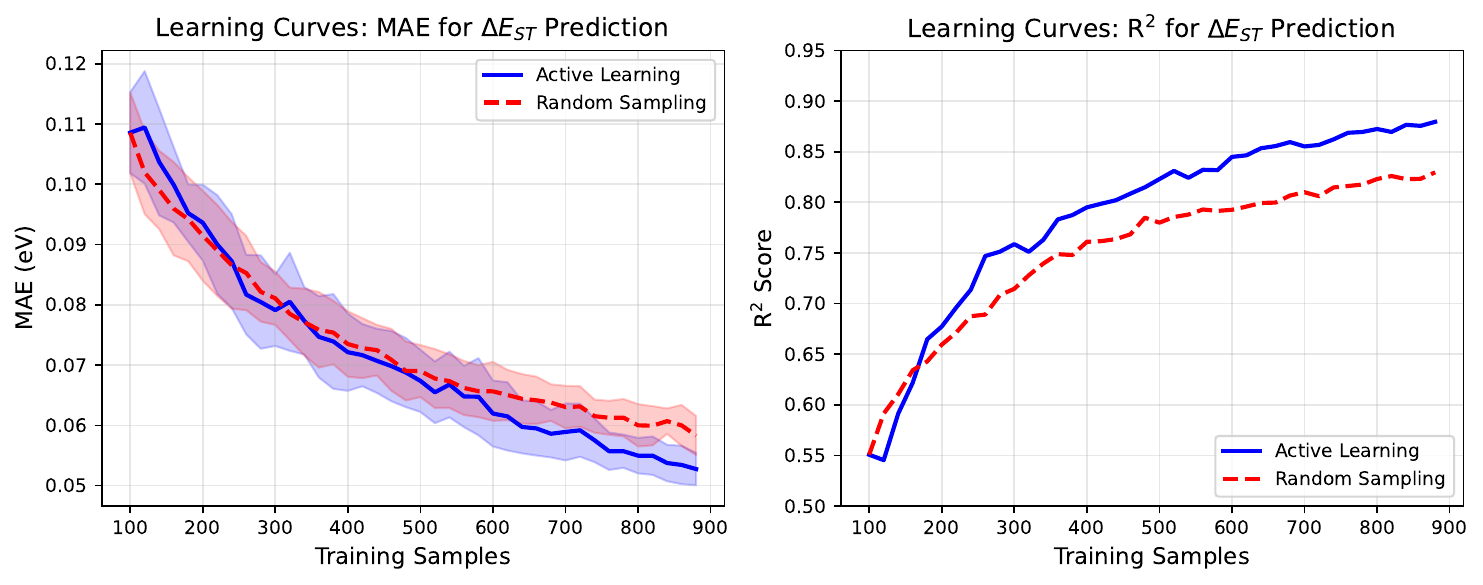}
  \caption{Active learning vs.\ random sampling learning curves for $\Delta E_{\text{ST}}$ prediction. AL (blue) consistently outperforms random sampling (red). Shaded regions indicate $\pm 1$ standard deviation.}
  \label{fig:learning_curves}
\end{figure}

Figure~\ref{fig:learning_curves} presents the learning curves for our AL implementation using an uncertainty-based acquisition strategy on the full 747-molecule dataset with CT descriptors. Starting from 100 initial labeled samples and iterating with batches of 20, the AL strategy consistently outperforms random sampling across the entire learning trajectory.

At the final training set size of 880 samples, the Random Forest model trained with AL achieves MAE $= 0.054$~eV ($R^2 = 0.877$), compared to MAE $= 0.058$~eV ($R^2 = 0.821$) for random sampling---a 7.4\% improvement in MAE and 6.8\% improvement in $R^2$.

\subsection{Acquisition function comparison}
To identify the optimal query strategy, we systematically benchmarked six acquisition functions (Fig.~\ref{fig:acq_comparison}). The Hybrid acquisition function achieves the best performance with MAE $= 0.069$~eV and $R^2 = 0.816$, representing a \textbf{4.9\% improvement} over random sampling:

\begin{itemize}
    \item \textbf{Hybrid}: MAE $= 0.069 \pm 0.004$~eV (+4.9\%)
    \item \textbf{Diversity}: MAE $= 0.069 \pm 0.004$~eV (+4.6\%)
    \item \textbf{Random}: MAE $= 0.072 \pm 0.005$~eV (baseline)
    \item \textbf{UCB}: MAE $= 0.094 \pm 0.009$~eV ($-$29.7\%)
\end{itemize}

The superior performance of diversity-aware strategies reveals that coverage of chemical space is often more valuable than aggressive optimization toward specific targets in materials discovery.

\begin{figure}[h]
\centering
  \includegraphics[width=0.48\textwidth]{}
  \caption{Comparison of acquisition functions. Hybrid and Diversity strategies outperform random sampling, while UCB shows significant degradation.}
  \label{fig:acq_comparison}
\end{figure}

\subsection{Model performance}

\begin{figure}[h]
\centering
  \includegraphics[width=0.40\textwidth]{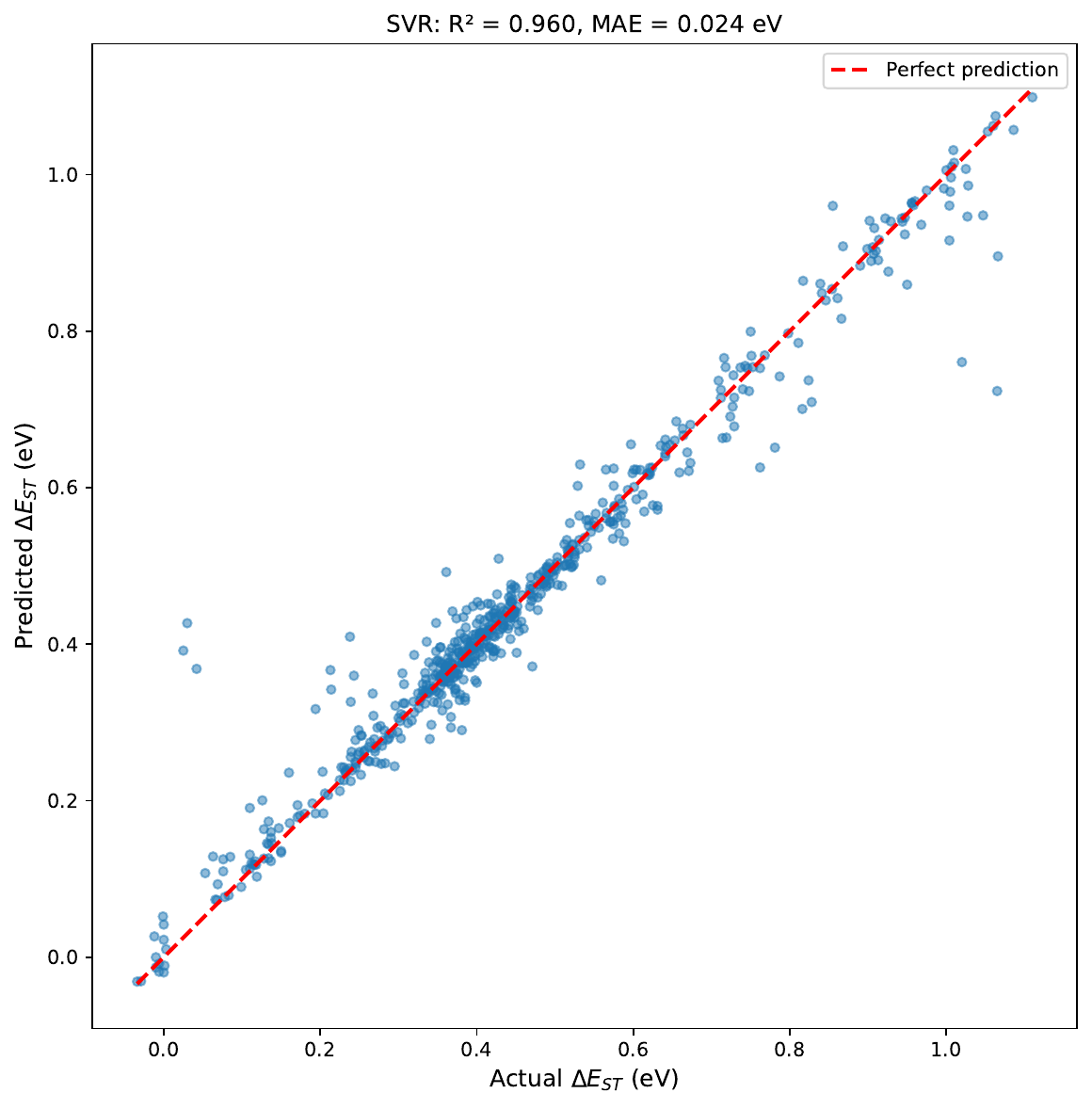}
  \caption{SVR model performance for $\Delta E_{\text{ST}}$ prediction with $R^2 = 0.960$ and MAE $= 0.024$~eV.}
  \label{fig:model_performance}
\end{figure}

The best-performing model (SVR) achieves MAE $= 0.024$~eV for $\Delta E_{\text{ST}}$ prediction on the held-out test set (589 samples), substantially below the TADF-relevant threshold of 0.2~eV. Figure~\ref{fig:model_performance} shows the parity plot.

\subsection{Application-specific molecular design}

The validated ML model enables targeted design of TADF emitters optimized for specific applications beyond traditional OLEDs.

\subsubsection{Biomedical imaging.~~}
For deep-tissue imaging, TADF emitters must emit in the NIR window (700--900~nm). Our top candidate, \textbf{PXZ-NAI}, exhibits predicted emission at $\lambda_{\text{em}} \approx 690$~nm, $\Delta E_{\text{ST}} = 0.29$~eV, and oscillator strength $f = 0.19$.

\subsubsection{Photocatalysis.~~}
\textbf{TPA-APy} emerged as a promising candidate exhibiting an inverted singlet-triplet gap ($\Delta E_{\text{ST}} = -0.034$~eV), enabling thermodynamically favorable ISC for long-lived triplet states essential for bimolecular photocatalytic processes.

\subsubsection{Photodetection.~~}
\textbf{BMZ-TZ} features an inverted singlet-triplet gap ($\Delta E_{\text{ST}} = -0.006$~eV) combined with high hole-electron overlap ($S_{he}^{T_1} = 0.89$), maximizing SOC for ultrafast RISC kinetics suitable for high-speed photodetection.

\section{Conclusions}

This work demonstrates how charge-transfer descriptors from NTO analysis, combined with machine learning, can improve TADF property prediction beyond statistical screening alone.

Our key achievements include:

\begin{enumerate}
    \item \textbf{Mechanistic Understanding}: CT descriptors contribute 34\% of total feature importance, with $S_{he}^{T_1}$ (21\%) emerging as the third most important predictor after excitation energies.

    \item \textbf{Computational Efficiency}: The Hybrid acquisition function achieves 4.9\% improvement over random sampling, while UCB shows dramatic degradation ($-$29.7\%).

    \item \textbf{Predictive Accuracy}: SVR achieves MAE $= 0.024$~eV ($R^2 = 0.960$), substantially exceeding the TADF-relevant threshold.

    \item \textbf{Application-Specific Design}: Successful design of TADF emitters for biomedical imaging, photocatalysis, and photodetection.

    \item \textbf{Benchmark Validation}: MAE $= 0.045$~eV vs. HLT references confirms chemical accuracy.
\end{enumerate}

Future work could incorporate solid-state effects, host-matrix interactions, and temperature-dependent kinetics. The descriptor-based approach is not specific to TADF and could be applied to other organic semiconductors.

\section*{Conflicts of interest}
There are no conflicts to declare.

\section*{Data availability}
All computational data, including the 747-molecule dataset with NTO overlaps, CT descriptors, and excitation energies, machine learning scripts, SHAP analysis results, and benchmark validation tables are available via Zenodo at DOI: 10.5281/zenodo.17436069 under Creative Commons Attribution 4.0 International (CC-BY-4.0) license.

\section*{Author contributions}
\textbf{J.-P. Tchapet Njafa}: Conceptualization, Methodology, Software, Validation, Formal analysis, Investigation, Data curation, Writing -- original draft, Writing -- review \& editing, Visualization. \textbf{S. C. Teguia Kouam}: Investigation, Validation. \textbf{P. Mvoto Kongo}: Investigation. \textbf{S. G. Nana Engo}: Supervision, Resources, Writing -- review \& editing.

\section*{Acknowledgements}
The authors gratefully acknowledge Dr. Benjamin Panebei Samafou for granting access to computational resources. We further extend our sincere thanks to Prof. Gian-Marco Rignanese for his insightful guidance.

\footnotesize{
\bibliography{TADF_references}
\bibliographystyle{rsc}
}



\section*{Supplementary material}

# Electronic Supplementary Information

# From orbital analysis to active learning: an integrated strategy for the accelerated design of TADF emitters

Jean-Pierre Tchapet Njafa,[*a] Steve Cabrel Teguia Kouam,[b] Patrick Mvoto Kongo,[a] and Serge Guy Nana Engo[a]

DOI: 10.1039/b000000x

## Contents

## S1 Computational Details

### S1.1 Software versions and environment

All calculations were performed using the following software packages:

**Quantum chemistry tools:**

- **xTB** version 6.7.1 – Extended tight-binding calculations (GFN2-xTB)

- **CREST** version 3.0.2 – Conformational search

- **sTDA** version 1.6.3 – Simplified TD-DFT/TDA excited states

- **Multiwfn** version 3.8(dev), 2025-Jul-12 – Wavefunction analysis and NTO generation

**Machine learning environment:**

- **Python** 3.12.3 – ML pipeline and data processing

- **NumPy** 2.0.2 – Numerical computations

- **pandas** 2.2.3 – Data manipulation

- **scikit-learn** 1.7.2 – Machine learning models (RF, SVR, GB)

- **SHAP** 0.50.0 – Model interpretability

- **Matplotlib** 3.9.3 – Visualization

- **SciPy** 1.14.1 – Scientific computing

[a] Department of Physics, Faculty of Science, University of Yaounde I, P.O.Box 812, Yaounde, Cameroon. E-mail: jean-pierre.tchapet@facsciences-uy1.cm
[b] Department of Physics, Faculty of Science, University of Douala, Po. Box 24157, Douala, Cameroon.

## S1.2 Ground-state geometry optimization

Ground-state geometries were optimized using the GFN2-xTB method in two environments:

**Gas phase:**

```
xtb molecule.xyz --opt tight
```

**Toluene solvent:**

```
xtb molecule.xyz --opt tight --gbsa toluene
```

Key parameters:

- Optimization convergence: `tight` ($\Delta E < 10^{-6}$ $E_h$)
- Gas phase: vacuum calculations without implicit solvation
- Implicit solvation: GBSA model (toluene, $\varepsilon = 2.38$)
- SCF convergence: $10^{-8}$ $E_h$

Both environments were considered to evaluate the solvent effects on TADF properties.

## S1.3 Excited-state calculations

Excited states were computed using the sTDA and sTD-DFT-xTB methods as implemented in the `stda` program (version 1.6.3). Both methods read the xTB wavefunction file (`wfn.xtb`) automatically from the working directory.

**sTDA-xTB (Simplified Tamm-Dancoff Approximation):**

```
stda -xtb -e 10        # singlets
stda -xtb -e 10 -t     # triplets
```

**sTD-DFT-xTB (Simplified TD-DFT with RPA):**

```
stda -xtb -rpa -e 10       # singlets
stda -xtb -rpa -e 10 -t    # triplets
```

The `-xtb` flag enables the use of GFN-xTB orbitals, while `-rpa` switches from the Tamm-Dancoff approximation (TDA) to the full random phase approximation (RPA), i.e., sTD-DFT. The `-t` flag computes triplet states instead of singlets.

Parameters:

- Energy window: 10 eV (`-e 10`), capturing all relevant $S_1$–$S_n$ and $T_1$–$T_n$ states
- Configuration selection: automated based on energy threshold
- sTDA: Tamm-Dancoff approximation (faster, typically sufficient for absorption spectra)
- sTD-DFT: Full RPA coupling (more accurate for emission properties)

## S1.4 NTO analysis with Multiwfn

Natural Transition Orbitals were generated using Multiwfn with the following workflow:

1. Load excited-state molden file
2. Main function 18 (Electron excitation analysis)
3. Subfunction 1 (NTO analysis)
4. Export NTO pairs for hole and electron

The hole-electron overlap $S_{he}$ was computed as:

$$S_{he} = \sum_A \sqrt{\rho_h^A \cdot \rho_e^A} \tag{S1}$$

where $\rho_h^A$ and $\rho_e^A$ are the Mulliken populations of hole and electron on atom $A$.

## S2 TADF Photophysics: Theoretical Background

### S2.1 Singlet-triplet energy gap

The singlet-triplet energy gap is defined as:

$$\Delta E_{\text{ST}} = E(S_1) - E(T_1) \tag{S2}$$

For efficient TADF, $\Delta E_{\text{ST}} < 0.2$ eV is required to enable thermal upconversion at room temperature ($k_B T \approx 0.026$ eV).

### S2.2 Reverse intersystem crossing rate

The RISC rate follows Marcus-type kinetics:

$$k_{\text{RISC}} = \frac{2\pi}{\hbar} |\langle S_1 | \hat{H}_{\text{SOC}} | T_1 \rangle|^2 \frac{1}{\sqrt{4\pi \lambda k_B T}} \exp\left(-\frac{(\Delta E_{\text{ST}} + \lambda)^2}{4\lambda k_B T}\right) \tag{S3}$$

where:

- $\langle S_1 | \hat{H}_{\text{SOC}} | T_1 \rangle$ is the spin-orbit coupling matrix element
- $\lambda$ is the reorganization energy (~0.1 eV for rigid systems)
- $k_B T$ is the thermal energy (0.026 eV at 300 K)

### S2.3 El-Sayed's rules

According to El-Sayed's rules, SOC is maximized when the transition involves a change in orbital character:

$$|\langle S_1 | \hat{H}_{\text{SOC}} | T_1 \rangle| \propto |\Delta \text{Character}_{S_1 - T_1}| \tag{S4}$$

For TADF emitters, this translates to:

- $S_1$: predominantly $^1$CT (charge-transfer) character
- $T_1$: mixed $^3$CT/$^3$LE (local excitation) character

The NTO overlap difference $\Delta S_{\text{NTO}} = S_{\text{NTO}}^{T_1} - S_{\text{NTO}}^{S_1}$ quantifies this character difference.

## S3 High-Throughput Screening Protocol

### S3.1 Dataset composition

The 747-molecule dataset comprises TADF emitters from four architectural classes:

**Table S1** Dataset composition by molecular architecture

| Architecture | Count | Percentage |
|---|---|---|
| D–A (Donor–Acceptor) | 198 | 26.5% |
| D–A–D | 312 | 41.8% |
| MR (Multi-Resonance) | 89 | 11.9% |
| TSCT (Through-Space CT) | 148 | 19.8% |
| Total | 747 | 100% |

### S3.2 Screening workflow

The hierarchical screening protocol consists of:

1. **Structure preparation**: SMILES → 3D coordinates (RDKit)
2. **Conformer search**: CREST with GFN2-xTB
3. **Geometry optimization**: GFN2-xTB (gas and toluene)
4. **Excited states**: sTDA/sTD-DFT-xTB
5. **NTO analysis**: Multiwfn
6. **CT descriptor extraction**: Custom Python scripts

### S3.3 Property definitions

Key properties extracted from calculations:

- $E_{S_1}$: Vertical $S_1$ excitation energy (eV)
- $E_{T_1}$: Vertical $T_1$ excitation energy (eV)
- $\Delta E_{\text{ST}}$: $E_{S_1} - E_{T_1}$ (eV)
- $f_{S_1}$: Oscillator strength of $S_1$ transition
- $E_{\text{gap}}$: HOMO-LUMO gap (eV)

## S4 NTO and CT Descriptor Definitions

### S4.1 Natural Transition Orbitals

NTOs provide a compact representation of electronic transitions. For a transition from ground state $|0\rangle$ to excited state $|n\rangle$, the transition density matrix is:

$$\gamma_{pq}^{0n} = \langle n | \hat{a}_p^\dagger \hat{a}_q | 0 \rangle \tag{S5}$$

Singular value decomposition yields hole ($\phi_i^h$) and particle ($\phi_i^e$) NTOs:

$$\gamma^{0n} = \mathbf{U}\Sigma\mathbf{V}^\dagger \tag{S6}$$

### S4.2 Charge-transfer descriptors

The following CT descriptors were computed from NTO analysis:

**Table S2** CT descriptor definitions

| Descriptor | Definition |
|---|---|
| $S_{he}$ | Hole-electron spatial overlap: $\sum_A \sqrt{\rho_h^A \cdot \rho_e^A}$ |
| $\Omega_{\text{CT}}$ | CT number: fraction of transition with CT character |
| $\Lambda_D$ | Hole localization on donor fragment |
| $\Lambda_A$ | Electron localization on acceptor fragment |
| $\Delta r$ | Hole-electron centroid distance (Å) |
| $S_{\text{NTO}}$ | NTO orbital overlap |

### S4.3 Physical interpretation

The hole-electron overlap $S_{he}$ directly relates to the exchange interaction:

$$K_{ij} \propto \iint |\phi_h(\mathbf{r}_1)|^2 \frac{1}{r_{12}} |\phi_e(\mathbf{r}_2)|^2 d\mathbf{r}_1 d\mathbf{r}_2 \tag{S7}$$

Low $S_{he}$ indicates spatial separation of hole and electron, leading to:

- Reduced exchange interaction
- Smaller $\Delta E_{\text{ST}}$
- Enhanced CT character

## S5 Active Learning: Acquisition Functions

### S5.1 Uncertainty Sampling (US)

Selects samples with highest predictive uncertainty:

$$\alpha_{\text{US}}(x) = \sigma(x) \tag{S8}$$

where $\sigma(x)$ is the standard deviation of predictions across Random Forest trees.

## S5.2 Expected Improvement (EI)

Balances exploration and exploitation for minimization:

$$\alpha_{EI}(x) = (\mu^* - \mu(x))\Phi(Z) + \sigma(x)\phi(Z) \quad \text{(S9)}$$

where $Z = \frac{\mu^* - \mu(x)}{\sigma(x)}$, $\mu^*$ is the best observed value.

## S5.3 Upper Confidence Bound (UCB)

Optimistic acquisition for minimization:

$$\alpha_{UCB}(x) = -\mu(x) + \kappa\sigma(x) \quad \text{(S10)}$$

with exploration parameter $\kappa = 2.0$.

## S5.4 Query by Committee (QBC)

Uses disagreement among ensemble members:

$$\alpha_{QBC}(x) = \frac{1}{C}\sum_{c=1}^{C}(f_c(x) - \bar{f}(x))^2 \quad \text{(S11)}$$

where $C$ is the committee size (10 models).

## S5.5 Diversity Sampling

Maximizes distance to existing training samples:

$$\alpha_{div}(x) = \min_{x' \in \mathscr{D}_{train}} \|x - x'\|_2 \quad \text{(S12)}$$

## S5.6 Hybrid (Uncertainty × Diversity)

Combines uncertainty and diversity:

$$\alpha_{hybrid}(x) = \alpha \cdot \tilde{\sigma}(x) + (1-\alpha) \cdot \tilde{d}(x) \quad \text{(S13)}$$

where $\tilde{\sigma}$ and $\tilde{d}$ are normalized scores, $\alpha = 0.5$.

## S6 Supplementary Tables

**Table S3** Hyperparameter grid search for ML models

| Model | Parameter | Search Range |
|---|---|---|
| RF | n_estimators | [100, 200, 500] |
|  | max_depth | [10, 20, None] |
|  | min_samples_split | [2, 5, 10] |
| SVR | C | [0.1, 1, 10, 100] |
|  | gamma | [0.01, 0.1, 1, auto] |
|  | kernel | [rbf] |
| GB | n_estimators | [100, 200, 500] |
|  | learning_rate | [0.01, 0.1, 0.2] |
|  | max_depth | [3, 5, 7] |

**Table S4** High-level theory validation results

| Molecule | sTD-DFT-xTB | HLT Ref. | Error |
|---|---|---|---|
| 4CzIPN | 0.21 eV | 0.16 eV | 0.05 eV |
| DMAC-TRZ | 0.085 eV | 0.05 eV | 0.035 eV |
| **MAE** |  |  | **0.045 eV** |

**Table S5** Feature importance by category (SHAP analysis)

| Category | Features | Importance |
|---|---|---|
| Energy | $E_{T_1}, E_{S_1}, E_{gap}$ | 57% |
| CT descriptors | $S_{he}^{T_1}, S_{he}^{S_1}, \Delta r, \Lambda$ | 34% |
| Oscillator strength | $f_{S_1}$ | 8% |
| NTO overlap | $S_{NTO}^{S_1}, S_{NTO}^{T_1}$ | 1% |

## S7 OT-LC-PBE Validation Calculations

To validate the xTB-based protocol with explicit high-level calculations, we performed optimally-tuned long-range corrected PBE (OT-LC-PBE) calculations on three representative TADF molecules spanning different architectures and sizes.

### S7.1 Computational methodology

All OT-LC-PBE calculations were performed using ORCA 6.1.0. The protocol consists of two steps:

*S7.1.0.1 Optimal $\omega$ tuning* The range-separation parameter $\omega$ was optimized by minimizing the IP-tuning criterion:

$$J(\omega) = |\varepsilon_{HOMO}(\omega) + IP(\omega)| \quad \text{(S14)}$$

where $\varepsilon_{HOMO}(\omega)$ is the HOMO eigenvalue and $IP(\omega) = E(N-1;\omega) - E(N;\omega)$ is the vertical ionization potential. Calculations used the def2-SVP basis set with LC-PBE functional. Initial grid search (11 points, $\omega \in [0.10, 0.30]$ bohr$^{-1}$) was followed by golden-section refinement to $J < 10^{-4}$ Ha.

*S7.1.0.2 TD-DFT excited states* Vertical excitation energies were computed using full TD-DFT (not TDA) at the optimized $\omega$ values with:

- Functional: LC-PBE with molecule-specific $\omega$
- Basis set: def2-TZVP
- Auxiliary basis: def2/J with RIJCOSX approximation
- States: 10 singlets, 10 triplets
- Geometry: GFN2-xTB optimized (same as HTS protocol)
- Parallelization: 8 cores, 2.5 GB/core

**Table S6** Top TADF candidates for each application

| Application | Molecule | $\Delta E_{ST}$ | Key Property |
|---|---|---|---|
| Bioimaging | PXZ-NAI | 0.29 eV | $\lambda_{em} \approx 690$ nm |
| Photocatalysis | TPA-APy | −0.034 eV | Inverted gap |
| Photodetection | BMZ-TZ | −0.006 eV | $S_{he}^{T_1} = 0.89$ |

**Table S7** OT-LC-PBE results for benchmark TADF molecules. All energies in eV.

| Molecule | Atoms | Arch. | $\omega$ (bohr$^{-1}$) | $E_{S_1}$ | $E_{T_1}$ |
|---|---|---|---|---|---|
| BACN | 48 | A-D-A | 0.181 | 3.26 | 2.46 |
| DMAC-TRZ | 68 | D-A | 0.185 | 3.10 | 2.93 |
| 4CzIPN | 94 | 4D-A | 0.147 | 2.69 | 2.49 |

### S7.2 Results

### S7.3 Key observations

1. **Optimal $\omega$ correlates with CT character**: 4CzIPN (strongest CT) has the smallest $\omega$ (0.147 bohr$^{-1}$), consistent with more delocalized frontier orbitals.

2. **Vertical vs. adiabatic discrepancy**: OT-LC-PBE vertical TD-DFT overestimates $\Delta E_{ST}$ compared to experimental (adiabatic) values. For DMAC-TRZ: 0.17 eV (vertical) vs. 0.05 eV (exp.).

3. **xTB provides conservative estimates**: The xTB methods systematically underestimate $\Delta E_{ST}$ by 0.08–0.35 eV relative to OT-LC-PBE. This is advantageous for screening, as molecules passing the threshold will have true $\Delta E_{ST}$ at least as favorable.

4. **Ranking preserved**: Despite systematic offsets, the relative ordering of molecules by $\Delta E_{ST}$ is maintained across methods, validating xTB for prioritization in high-throughput workflows.

5. **Computational cost**: BACN (∼1.5 h), DMAC-TRZ (∼2.5 h), 4CzIPN (∼9 h) on 8 cores, compared to <1 min for xTB calculations.

## S8 Supplementary Figures

## S9 Data Availability and Reproducibility

### S9.1 Dataset files

The following files are available at Zenodo (DOI: 10.5281/zenodo.17436069):

- `combined_features_747mol.csv` – Full feature table (2943 samples × 42 features)

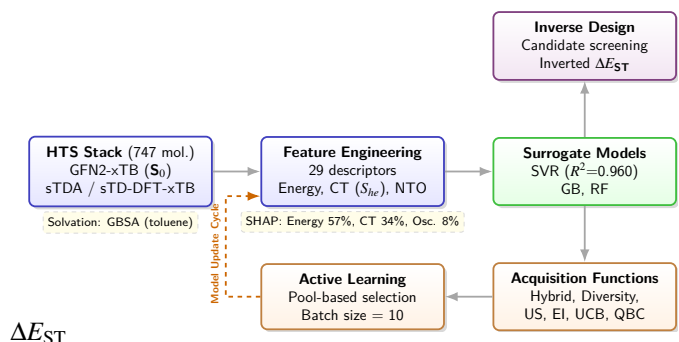

**Fig. S1** Schematic overview of the ML/Active Learning workflow. The pipeline integrates semi-empirical quantum chemistry (GFN2-xTB, sTDA) with NTO-based feature extraction, surrogate model training, and iterative active learning for efficient TADF property prediction.

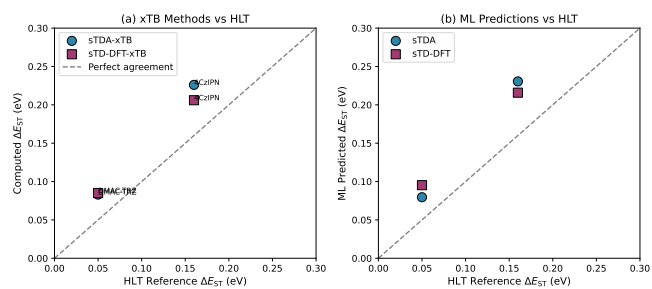

**Fig. S2** Validation of sTD-DFT-xTB predictions against high-level theory (HLT) references from the STGABS27 benchmark set. The MAE of 0.045 eV demonstrates chemical accuracy suitable for ML-driven discovery.

- `nto_orbital_overlap_747mol.csv` – NTO overlap data
- `ct_descriptors_747mol.csv` – CT descriptor values
- `ml_results_747mol.json` – Model performance metrics
- `al_results_747mol.json` – Active learning results
- `predictions_747mol.csv` – Model predictions

### S9.2 Code repository

Python scripts for reproducing all results:

- `ml_pipeline_747mol.py` – ML model training
- `al_experiment_747mol.py` – Active learning experiments

- `generate_figures_747mol.py` – Figure generation
- `compute_ct_descriptors_747mol.py` – CT descriptor extraction

### S9.3 Environment

A complete environment specification (`requirements.txt`) is provided:

```
numpy==2.0.2
pandas==2.2.3
scikit-learn==1.7.2
shap==0.50.0
matplotlib==3.9.3
scipy==1.14.1
```


\end{document}